\documentstyle[12pt,aps,oldlfont]{revtex}

\def\NP#1{Nucl.\ Phys.\ {\bf  #1}}
\def\PR#1{Phys.\ Rev.\ {\bf  #1}}

\def\PL#1{Phys.\ Lett.\ {\bf  #1}}

\def\half{{\textstyle{1\over 2}}}

\def\psibar{\overline\psi}
\def\qbar{\overline q}

\def\etal{{\it et al.}}

\newcommand{\tr}{{\rm tr}\;}                                    % trace !
\newcommand{\wt}[1]{\widetilde{#1}}                             % Wide tilde
\newcommand{\sla}[1]{\makebox[1pt][l]{$#1$}/ \;}                % Slash!

\newcommand{\be}{\begin{equation}}                              % Reduces
\newcommand{\ee}{\end{equation}}                                % typing!
\newcommand{\bea}{\begin{eqnarray}}
\newcommand{\eea}{\end{eqnarray}}
\newcommand{\nn}{\nonumber}
\newcommand{\rar}{\rightarrow}

\newcommand{\bra}{\langle}
\newcommand{\ket}{\rangle}

\begin{document}
\title{A nonlocal, covariant generalisation of the NJL model}
\author{R. D. Bowler and M. C. Birse}
\address{Theoretical Physics Group\\
Department of Physics and Astronomy\\
The University of Manchester\\
Manchester, M13 9PL, UK\\}
\maketitle
\vspace{10pt}
\begin{abstract}
We solve a nonlocal generalisation of the NJL model in the Hartree
approximation. This model has a separable interaction, as suggested by
instanton models of the QCD vacuum. The choice of form factor in this
interaction is motivated by the confining nature of the vacuum. A conserved
axial current is constructed in the chiral limit of the model and the pion
properties are shown to satisfy the Gell-Mann--Oakes--Renner relation. For
reasonable values of the parameters the model exhibits quark confinement. \\
\end{abstract}

\section{Introduction}
The Nambu--Jona-Lasinio (NJL) model \cite{njl} is often used as a model to
study the hidden chiral symmetry of QCD \cite{njlrev}. It describes fermions
interacting via a local, chirally invariant four-point coupling. For large
enough coupling strengths it leads to a vacuum state with a quark condensate
and pions which are approximate Goldstone bosons. The local nature of the
interaction considerably simplifies the calculations compared with ones based
on
finite-range forces. However it also means that loop diagrams are divergent and
must be regularised in some way.

This regularisation scheme forms part of the model and introduces a cut-off as
an extra parameter. Many procedures have been used in the literature, for
example: three- and four-momentum cut-offs, Pauli--Villars, proper-time
regularisation \cite{njlcut}. All of these yield qualitatively similar results.
None however has any clear interpretation in terms of the underlying theory,
QCD. Cutting off the quark propagator at high momenta, as most of these schemes
do, does violence to the Hilbert space. In loop diagrams with two internal
quark propagators, such as are needed for calculating mesonic bound states, one
has to be careful about how the cut-off is imposed on both propagators if axial
Ward identities are not to be violated. Furthermore, while a cut-off is needed
to make the nonanomalous part of the effective action finite, cutting off the
anomalous part means that, for example, the amplitude for $\pi^0\rar 2\gamma$
is not given correctly.\footnote{For a review of these problems, see
\cite{ballrip}.}

An attractive alternative to the usual local NJL model is suggested by the
instanton picture of the QCD vacuum studied by Diakonov and Petrov
\cite{dp1,dp2}. The interactions of the quarks with the instantons generate an
effective four-point coupling (in the case of two light quark flavours). The
nonlocality of this interaction provides a natural cut-off on the loop
integrals. Moreover the nonlocality a separable form. This considerably
simplifies the Schwinger--Dyson and Bethe--Salpeter equations in comparison
with
models based on quark-quark forces motivated by gluon exchange.\footnote{See
\cite{hrw} for a recent example of such a model and many further references.}

In Diakonov and Petrov's original version the form factor in this interaction
is given by the zero mode of a quark in the presence of an instanton. Here we
consider more general choices, motivated in part by the fact that the confining
nature of the QCD vacuum will modify the long-range behaviour of the form
factor. We explicitly construct a conserved axial current for this type of
model and demonstrate that PCAC, implicitly assumed by Diakonov and Petrov,
does indeed hold. In addition we show that a rather natural choice of form
factor can lead to quark confinement. We also compare our model with other
nonlocal NJL-like models based on separable interactions.

\section{Nonlocal model}
The action for a generalised NJL model with a nonlocal four-point interaction
and approximate SU(2)$\times$SU(2) chiral
symmetry is
\bea
S&=&\int d^4\!x\;\psibar(x) \big(i \sla{\partial} -m_c\big)\psi(x) \\
& &\mbox{}+\int d^4\!x_1\ldots d^4\!x_4\;\alpha (x_1,x_2,x_3,x_4)\nn\\
& &\qquad\qquad\qquad\qquad\mbox{}\times\bigg[\psibar(x_1)\psi(x_3)
\psibar(x_2)\psi(x_4)+\psibar(x_1) i\gamma_5\tau^a \psi(x_3)
\psibar(x_2) i\gamma_5\tau^a \psi(x_4) \bigg]. \nn
\eea
The symmetry is explicitly broken by the current quark mass $m_c$. For a
separable interaction the smearing function may be written in momentum space as
\be\wt{\alpha}(p_1,p_2,p_3,p_4)=\half(2\pi)^8Gf(p_1)f(p_2)f(p_3)f(p_4)\;\delta
(p_1+p_2-p_3-p_4), \ee
where we take the form factors $f(p)$ to be normalised such that $f(0)=1$.
In the limit of a local interaction, $f(p)=1$ for all $p$, this reduces to
the ordinary NJL model with coupling strength $G$.

As in the ordinary NJL model, for a sufficiently large coupling the quarks
acquire a dynamical mass. With a nonlocal interaction this dynamical mass
$M(p)$ is momentum-dependent. It can be determined, in the Hartree
approximation, from the gap equation:
\be M(p)=m_c+G \nu f(p)^2\int \frac{d^4\!k}{(2\pi)^{4}}
\frac{M(k)f^2(k)}{k^2+M(k)^2},
\ee
where $\nu = 4 N_c N_f$, $N_c$ and $N_f$ being the numbers of colours and
flavours respectively. The momentum integral has been Wick rotated so that the
four-momentum $k$ is in Euclidean space. From now on all momenta will be
Euclidean, unless explicitly stated otherwise. The separable nature of our
interaction means that the loop integral is independent of the external
momentum $p$ and so this equation can be solved in exactly the same way as in
the NJL model.

In choosing the form factor $f(p)$, we obviously want it to be Lorentz
invariant. We assume that the interaction is generated by nonperturbative
features of the QCD vacuum and so does not have the divergent short-distance
behaviour of one-gluon exchange. Hence we want a form that falls off
sufficiently fast at high momenta to keep the loop integrals finite. At low
momenta we want it to be more regular than the instanton form used by Diakonov
and Petrov since in coordinate space it corresponds to a function that falls
off rapidly with distance. Such a fall-off is to be expected from the confining
nature of the QCD vacuum. The simplest choice satisfying all of these
requirements is a Gaussian function of the Euclidean four-momentum:
\be
f(p)=\exp(-p^2/\Lambda^2),
\ee
where $\Lambda$ is our cut-off parameter describing the range of the
nonlocality. We use this form factor in the numerical calculations presented
here. Other choices which also possess these features are likely to lead to
qualitatively similar results.

With this choice of form factor, the model does exhibit confinement of the
quarks for large enough values of the dynamical mass. This arises because the
equation $p^2=M(p)^2=M(0)^2\exp[2p^2/\Lambda^2]$ in Minkowski space has no real
solutions. Hence the quark propagator has no real poles and quarks do not
appear as asymptotic states. The quark propagator does have a pair of poles
with complex masses, corresponding to quarks which have a finite lifetime as
isolated particles. This is similar to the situation in other models for
confinement based on approximate Schwinger-Dyson equations \cite{conf,hrw}.

The model has three adjustable parameters: $G$, $\Lambda$ and $m_c$. We fix two
of these using the pion mass and decay constant. The remaining free parameter
can then be chosen to give reasonable values for the dynamical quark mass and
decay constant.

\section{Bethe-Salpeter equation}
The pion is constructed, as in the NJL model, by solving the corresponding
Bethe-Salpeter equation. For a separable interaction this equation is just a
geometric series. For example, the $T$-matrix in the pseudoscalar isovector
channel can be written in the form
\bea
T(p_1,p_2,p_3,p_4)&=& f(p_1)f(p_2)f(p_3)f(p_4)\big[G + G^2 J_{\rm PP}(p)+
G^3 J_{\rm PP}(p)^2+ \cdots\big]\\
&=& f(p_1)f(p_2)f(p_3)f(p_4){G\over 1-G J_{\rm PP}(p)},\nn
\eea
where $p=p_1+p_2=p_3+p_4$ and $J_{\rm PP}$ is given by a quark loop with two
nonlocal pseudoscalar insertions:
\be
J_{\rm PP}(p) = \tr \int \frac{d^4\!k}{(2\pi)^{4}}
\left[i \gamma^5
\frac{f^2(k+\frac{1}{2}p)}{\sla{k}+\frac{1}{2}\sla{p}+M(k+\frac{1}{2}p)}
i \gamma^5 \frac{f^2(k-\frac{1}{2}p)}{\sla{k}-\frac{1}{2}
\sla{p}+M(k-\frac{1}{2}p)} \right]\;.
\ee
The pion mass and wave function renormalisation can be determined from the
position and residue of the pole in the propagator in this channel. For momenta
in the vicinity of the pion pole, $p^2=-m_\pi^2$, we write
\be
{G\over 1-GJ_{\rm PP}(p)}\equiv{Z_\pi\over p^2+m_\pi^2}+\cdots.
\ee

Assuming that the symmetry-breaking current quark mass is much smaller than the
dynamical mass, we now expand $J_{\rm PP}$ about the chiral limit. To first
order in the current quark mass and $p^2$ it can be written in the simple form
\be
J_{\rm PP}(p)\simeq\frac{1}{G} -\frac{m_c}{M_0(0)} \bra \psibar \psi \ket
- \frac{p^2}{Z_{\pi}},
\ee
where $M_0(p)$ is the dynamical mass at zero current quark mass. In obtaining
this we have made use of the gap equation and the expression for the quark
condensate,
\be
\bra \psibar \psi \ket =\nu \int \frac{d^4\!k}{(2\pi)^{4}}
\frac{M_0(k)}{k^2+M_0(k)^2},
\ee
in the chiral limit. To leading order in $m_c$, the pion mass is given by
\be
m_\pi^2=-{Z_\pi\over M_0(0)^2}m_c\bra \psibar \psi \ket,
\ee
and the wave function renormalisation by
\be
Z_{\pi}^{-1}=\frac{\nu}{2M_0(0)^2} \int \frac{d^4\!k}{(2\pi)^{4}}
\frac{M(k)^2-M'(k) M(k) k^2 + \big(M'(k)\big)^2 k^4}
{\big(k^2+M_0(k)^2\big)^2},
\ee
where the prime denotes differentiation with respect to $k^2$.

\section{Pion decay constant}
In order to calculate the pion decay constant $F_{\pi}$ for this model we need
to construct an axial current which is conserved in the chiral limit. The usual
local current,
\be
j^{\mu a}_{5({\rm loc})}=\half\psibar\gamma^\mu\gamma_5\tau^a\psi,
\ee
is not conserved in the presence of the nonlocal interaction of Eq.~(1) and so
it cannot be the symmetry current of this model. This can be seen from the
results of Diakonov and Petrov \cite{dp2} where the pion decay constant
calculated from this current fails to satisfy the Gell-Mann--Oakes--Renner
(GOR) relation \cite{gmor}. In addition, the correlator for this current is not
transverse, another signal that it is not the appropriate symmetry current.

Various techniques are available for construction of a conserved current in the
presence of nonlocal interactions. A rather elegant one is to make the action
locally invariant by coupling the quarks to right- and left-handed gauge
fields, $A^\mu_{R,L}$ \cite{bkn}. Some of these fields can be identified with
those of the electromagnetic or weak interactions, but in general they need not
be physical. The corresponding currents are obtained by differentiating the
action with respect to the gauge fields and then setting those fields equal to
zero.

The locally invariant action is constructed by replacing the ordinary
derivative by the covariant one in the kinetic term. The nonlocal interaction
can be made invariant by introducing parallel transport operators,
\be
W_{R,L}(x,y) = {\rm Pexp} \left[ \int_0^1 \!d\lambda\, {dz^\mu\over d\lambda}
A_{\mu}(z)(1\pm\gamma_5)\right],
\ee
where the integral follows a path $z(\lambda)$ between the appropriate right-
or left-handed quark fields at space-time points $x$ and $y$. This is analogous
to the gauge-invariant point splitting used in deriving anomalies
\cite{jackiw}. Although this method is convenient in a version of the model (1)
with only one quark flavour, it becomes rather cumbersome for the in a model
with two or more flavours because of the non-abelian nature of the chiral
symmetry.

Alternatively one can use a Noether-like method of construction, evaluating the
divergence of the local current with the aid of the equations of motion
obtained from the action (1) (with $m_c=0$). For the axial isospin current,
which we are interested in here, this gives
\bea
\partial_{\mu}\left(\half\psibar\gamma^\mu\gamma_5\tau^a\psi\right)&=&
-\int d^4\!x_1\ldots d^4\!x_4\, \alpha(x_1,x_2,x_3,x_4)
  \psibar(x_1) \tau_a i\gamma^{5} \psi(x_3)\psibar(x_2) \psi(x_4)\\
& &\qquad\qquad\mbox{}\times\left[ \delta(x_1-x)
   -\delta(x_2-x)
   +\delta(x_3-x)-\delta(x_4-x)\right]\nn\\
& &\mbox{}+\half\int d^4\!x_1\,\ldots d^4\!x_4\, \alpha(x_1,x_2,x_3,x_4)
   \psibar(x_1)[\tau_a,\tau_b]\psi(x_3) \psibar(x_2)\tau_b i\gamma^{5}\psi(x_4)
\nn\\
& &\qquad\qquad\mbox{}\times\left[\delta(x_1-x)
   -\delta(x_3-x) \right] ,\nn
\eea
where we have assumed that $\alpha$ is symmetric under permutations of its
arguments. The r.h.s.~of this equation, which arises from the variation of the
nonlocal interaction under local axial isospin transformations, must be
re-expressed as a divergence if we are to identify a conserved current. Ball
and Ripka \cite{ballrip} do this by making a derivative expansion of the
nonlocality.

However such an expansion is not essential: a trick suggested by the previous
method is to write the differences of delta functions as line integrals, for
example,
\be
\delta(x_1-x)-\delta(x_4-x)=\int_0^1 d\lambda\,{dz^\mu\over d\lambda}
{\partial\over \partial x^\mu}\delta(z-x),
\ee
where the path $z(\lambda)$ runs from $x_1$ to $x_4$. Interchanging the order
of differentiation with respect to $x$ and the various integrals leaves the
r.h.s.~of (14) in the form of a total divergence from which the the nonlocal
contribution to the symmetry current can be identified. If we choose, for
convenience, straight line paths such as
\be
z(\lambda)=(1-\lambda)x_1+\lambda x_4,
\ee
then the nonlocal current is
\bea
j^{\mu a}_{5({\rm nl})} &=& \int d^4\!x_1\ldots d^4\!x_4\,
\int^1_0 d\lambda\,\alpha(x_1,x_2,x_3,x_4)\\
& &\mbox{}\times\Big\{\psibar(x_1) \tau_a i\gamma^5 \psi(x_3)\psibar(x_2)
\psi(x_4)\Big[(x_4^{\mu}-x_1^{\mu})
\delta\Big(\lambda x_1+(1-\lambda)x_4-x\Big)\nn\\
& &\qquad\qquad\qquad\qquad\qquad\qquad\qquad\qquad
\mbox{}-(x_3^{\mu}-x_2^{\mu})
\delta\Big(\lambda x_2+(1-\lambda)x_3)-x\Big)\Big]\nn\\
& &\qquad\mbox{}-i\epsilon^{abc}\psibar(x_1)\tau_c\psi(x_3)
\psibar(x_2)\tau_b i\gamma^5\psi(x_4)
(x_3^{\mu}-x_1^{\mu})\delta\Big(\lambda x_1+(1-\lambda)x_3-x\Big)\Big\}.\nn
\eea
The full current is the sum of this and the local piece (12).

Note that none of these procedures determines the current uniquely. The
ambiguity shows up most clearly in the arbitrary choice of path for the line
integrals but it affects all methods for constructing a conserved current from
a nonlocal action. It arises because the requirement that the divergence of the
current vanish only fixes the longitudinal part of the current; the transverse
part remains undetermined. In the derivative expansion of \cite{ballrip} it
corresponds to the fact that one can always add to the current combinations of
field derivatives whose divergence vanishes by construction. This problem is
well known in nuclear physics: the longitudinal components of exchange currents
can be related to phenomenological nucleon-nucleon forces, while the transverse
currents require a specific model for the underlying meson
exchanges.\footnote{See, for example, \cite{mec}.} For our present purpose,
calculation of $F_\pi$, only the longitudinal current is needed and so our
results are not affected by this ambiguity.

The pion decay constant, $-i F_{\pi}p^{\mu}=\bra 0|j^{\mu a}_5(0)
|\pi^a(p)\ket$, can be found by extracting the pion pole from the correlator of
two axial currents. This correlator can be written
\be
J_{\rm PA}^{\mu}(p)^\dagger {G\over 1-GJ_{\rm PP}(p)}J_{\rm PA}^{\nu}(p)
={4 F_\pi^2 p^\mu p^\nu\over p^2+m_\pi^2}+\cdots,
\ee
where $J_{\rm PP}(p)$ is given by the quark loop described in the previous
section  and $J_{\rm PA}^{\mu}(p)$ is given by a similar loop with one
pseudoscalar and one axial current insertion. Equating the residues at the
poles on  both sides, one finds
\be
4 F_{\pi}^2p^{\mu}p^\nu=Z_\pi  J_{PA}^{\mu}(p)^\dagger J_{PA}^{\nu}(p),
\ee
where $p^2=-m_\pi^2$.

The contribution of the local piece of the current to $J_{\rm PA}^{\mu}(p)$ is
straightforward and has the form:
\be
J_{\rm PA(loc)}^\mu(p) = \tr \int \frac{d^4\!k}{(2\pi)^{4}}
\left[\gamma^\mu\gamma^5
\frac{1}{\sla{k}+\frac{1}{2}\sla{p}+M(k+\frac{1}{2}p)}
i \gamma^5 \frac{f(k+\frac{1}{2}p)f(k-\frac{1}{2}p)}{\sla{k}-\frac{1}{2}
\sla{p}+M(k-\frac{1}{2}p)} \right]\;.
\ee
The nonlocal contribution is more involved and so we do not present the details
of its evaluation here. Its main features can be seen from the
form of the the nonlocal current (17) in momentum space,
\bea
\tilde j^{\mu a}_{5({\rm nl})}(p) &=&   G \frac{p^{\mu}}{p^2}
\int \frac{d^4\!k_1\ldots d^4\!k_4}{(2\pi)^{8}}
\psibar(k_1) i\gamma_5 \tau^a \psi(k_3) \psibar(k_2) \psi(k_4)
\delta(k_1+k_2-k_3-k_4+p)\\
& & \qquad\times
\Big[f(k_1+p)f(k_2)f(k_3)f(k_4)-f(k_1)f(k_2+p)f(k_3)f(k_4)\nn\\
& & \qquad\qquad+f(k_1)f(k_2)f(k_3+p)f(k_4)-f(k_1+p)f(k_2)f(k_3)f(k_4+p)
\Big]+\cdots,
\nn\eea
where the terms involving commutators of isospin matrices have not been
written out since they do not contribute to $F_\pi$. Each of the terms shown
in Eq.~(21) involves the product of pseudoscalar isovector and scalar isoscalar
densities. The vacuum to one-pion matrix element of the nonlocal current
consists of similar terms, each of which is a product of two integrals: one
similar in its Dirac structure to $J_{\rm PP}$, the other like the self-energy
in Eq.~(3).

After a certain amount of effort, the total $J_{\rm PA}^\mu(p)$ can be
expressed in terms of the same integral as in $Z_\pi^{-1}$ above, and so
can be written as
\be
J_{\rm PA}^\mu(p)=-2ip^\mu M_0(0)Z_\pi^{-1}
\ee
to lowest order in $m_c$. Using this in Eq.~(19), we see that
$F_\pi$ can be related to the wave function renormalisation by
\be
F_\pi^2=M_0(0)^2Z_\pi^{-1}.
\ee
The GOR relation follows at once from this and Eq.~(10) for the pion mass:
\be
F_{\pi}^2 m_{\pi}^2 = -m_c \bra \psibar \psi \ket.
\ee

\section{Results and discussion}
We fix the current quark mass $m_c$ and one combination of $G$ and $\Lambda$ to
give a pion mass of 140 MeV and a pion decay constant of 93 MeV. Results for
the dynamical quark mass and the quark condensate are shown in the table. The
variation of these quantities with the cut-off $\Lambda$ is qualitatively
similar to that obtained using the local NJL model with the usual
regularisation schemes \cite{njlrev,njlcut}.

Typically the dynamical quark mass in quark models is taken to be in the
region 300--400 MeV. The average current mass of the up and down quarks is
believed to be about 5--10 MeV at the momentum scale relevant to hadronic
models \cite{qmass}. From the GOR relation, this corresponds to a quark
condensate in the range $\langle\qbar q\rangle \equiv
\half\langle\psibar\psi\rangle \simeq -(210\ \hbox{MeV})^3$ to $-(260\
\hbox{MeV})^3$. From the table we see that parameter sets with cut-offs of
around 1 GeV yield reasonable results for quark masses and the condensate.

For large enough values of the dynamical mass, the quark propagator ceases to
have poles at real $p^2$ in Minkowski space, and quarks become confined.
In the chiral limit, this happens for
\be
{M_0(0)\over\Lambda}>{\hbox{e}^{-{1\over 2}}\over 2}.
\ee
Neglecting the effect of the small current quark mass, we see that our quarks
are confined for those parameter sets which give $M_0(0)$ above about 300 MeV.
The critical point for confinement is thus in the middle of the region which
yields reasonable values for the condensate.

Other related models have been studied by various authors. In particular
Diakonov and Petrov \cite{dp1} have used a nonlocal interaction with a
separable form suggested by an instanton approach to the QCD vacuum. The
expression for $F_\pi$ given by Eqs.~(23, 11) agrees with theirs \cite{dp2}.
However those authors never explicitly calculate $F_\pi$ from the conserved
axial current. Instead they take the pion wave function renormalisation, and
assuming that PCAC holds, determine $F_\pi$ from it using Eq.~(23) (the
equivalent is Eq.~(30) of their paper). When they attempt to calculate $F_\pi$
using the local piece of the current alone, they find an expression which was
originally obtained by Pagels and Stokar \cite{pagsto}. The incompleteness of
this expression of Pagels and Stokar and the fact that the full symmetry
current requires a nonlocal contribution have also been pointed out by Holdom
and coworkers in the context of a constituent quark model with a nonlocal
self-energy\cite{holdom}. In fact the model of \cite{holdom} has many features
in common with the one we study here, although it is not derived from an
underlying quark-quark interaction. Ball and Ripka \cite{ballrip} discuss in
detail the construction of conserved currents in such models.

Buballa and Krewald \cite{bubkre} have also studied a model of the form (1),
with a different choice of form factor. They too find quark confinement but
their form factor is chosen to avoid the appearance of complex poles in the
quark propagator. The propagator in their model does however contain some
unusual analytic properties, in the form of additional cuts arising from the
square roots in the form factor. The appearance of extra singularities in the
quark propagator seems unavoidable in models for confinement based on
Schwinger-Dyson equations \cite{conf,hrw}. Moreover the model of
Ref.~\cite{bubkre} still requires an explicit cut-off since the form factor
fails to regulate the momentum integrals in loops.

Another related model has been studied by the Rostock group \cite{rostock}. It
is based on an instantaneous interaction with a separable dependence on the
three-momenta. That model is of course not covariant and so, for example, there
will be different pion decay constants associated with the space and time
components of the axial current. Schmidt \etal\ have calculated $F_\pi$ from
the
time component; since their interaction is instantaneous, that component of the
current is purely local and the extra terms described here do not appear.

\section*{Acknowledgements}
We are grateful to D. Blaschke, C. Christov, G. Ripka and S. Schmidt for
helpful discussions. We wish to thank the ECT$^*$, Trento and the University
of Rostock for their hospitality. MCB is supported by an EPSRC Advanced
Fellowship.

\section*{Table}

\begin{table}
\begin{center}
\begin{tabular}{rrrr}
\hline
$M_0(0)$&$\Lambda$&$-\langle\qbar q\rangle^{1\over 3}$&$m_c$\\
\hline
200&2222&330&2.4\\
250&1414&257&5.0\\
300&1075&222&7.7\\
350&887&200&10.5\\
400&768&185&13.3\\
450&684&173&16.2\\
500&623&165&18.9\\
\hline
\end{tabular}
\vspace{10pt}
\caption{Results for dynamical quark mass and quark condensate, with
$F_\pi=93$ MeV and $m_\pi=140$ MeV. All quantities are in MeV.}
\end{center}
\end{table}

\end{document}